\documentclass[aps,pre,amsmath,amssymb,twocolumn,showpacs,byrevtex,superscriptaddress]{revtex4}
\usepackage{graphicx}
\begin{document}
\title{Noro-Frenkel scaling in short-range square well:  A Potential Energy Landscape study}.
 
\author{Giuseppe~Foffi}\email{giuseppe.foffi@epfl.ch}
                 \affiliation{Institut Romand de Recherche Num\'erique en
                   Physique des Mat\'eriaux IRRMA,
                   PPH-Ecublens, CH-105 Lausanne, Suisse}         

\author{Francesco~Sciortino} \email{francesco.sciortino@phys.uniroma1.it}\affiliation{
         {Dipartimento di Fisica and INFM-CNR-CRS Soft, Universit\`a di Roma {\em La  Sapienza}, P.le A. Moro 2, 00185 Roma, Italy}
        }
        
\begin{abstract}
  We study the statistical properties of the potential energy landscape of a
  system of particles interacting via a very short-range square-well potential
  (of depth $-u_0$), as a function of the range of attraction $\Delta$ to
  provide thermodynamic insights of the Noro and Frenkel~[ M.G. Noro and D.
  Frenkel, J.Chem.Phys.  {\bf 113}, 2941 (2000)] scaling.  We exactly evaluate
  the basin free energy and show that it can be separated into a {\it
    vibrational} ($\Delta$-dependent) and a {\it floppy}
  ($\Delta$-independent) component.  We also show that the partition function
  is a function of $\Delta e^{\beta u_o}$, explaining the equivalence of the
  thermodynamics for systems characterized by the same second virial
  coefficient.  An outcome of our approach is the possibility of counting the
  number of floppy modes (and their entropy).
\end{abstract}


\maketitle

\begin{figure}[tbh!]
\includegraphics[width=.34\textwidth]{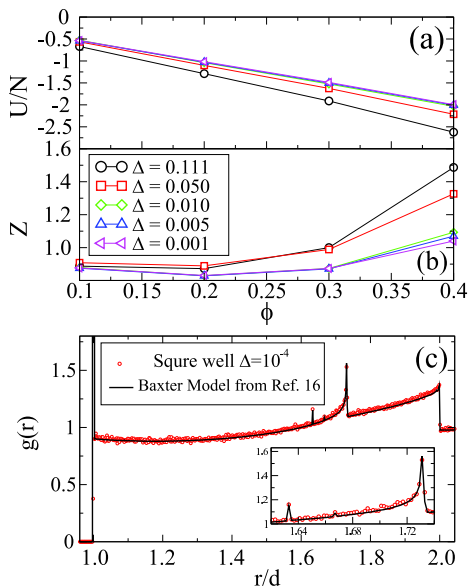}
\caption{(Color online) Top: (a) Energy per particle vs. $\phi$ for various $\Delta$ at
  $B_2^*=-0.40$. (b) same as (a) for compressibility factor $Z=PV/Nk_BT$. (c)
  Radial distribution function for $\phi=0.164$ and $B_2^*=-0.92$. The
  continuous line is from simulation of the Baxter model~\cite{Miller2004},
  dots are for a SW with $\Delta=10^{-4}$.  }
\label{fig:NF}
\end{figure}

Colloid and protein systems are often characterized by interactions on a
length scale significantly smaller than the particle size.  When the
interaction range is less than $10\%$ of the repulsive diameter, both
termodynamic and dynamic properties are drastically different from the
standard liquid behavior ~\cite{Anderson2002, Trappe2004,Dawson2001etal,
  Pham2002etal,Sciortino2005a}.
An important characteristic of short-range spherical potentials $V(r)$ is the
independence of thermodynamic properties by the specific $V(r)$ shape.  In
particular, the reduced second virial coefficient $B_2^* \equiv B_2/B_2^{HS}
\equiv -2 \pi \int \left( 1- e^{-\beta V(r)} \right) r^2 dr/B_2^{HS} $ (where
$B_2^{HS}$ is the hard-sphere $B_2$) appears to be a proper scaling variable
which combines temperature $T$ and potential shape. Based on the analysis of
several available theoretical and numerical data, Noro and Frenkel
(NF)~\cite{Noro2000} have proposed a generalized law of corresponding states
which states that all short-ranged spherically symmetric attractive potentials
are characterized by the same thermodynamics properties if compared at the
same reduced density and  $B_2^*$.  This universal behavior is
captured by the extreme limit of vanishing interaction range, i.e. by the
celebrated Baxter sticky sphere model~\cite{Baxter1968b}, which, despite its
known pathologies~\cite{Stell1991,Fishman1981} has been successfully used to
interpret experimental results in colloidal and proteins systems.  The
available accurate estimates of the critical point location and two-phase
coexistence of the Baxter model~\cite{Miller2003} are becoming the reference
for the thermodynamic behavior of all short-range spherical potentials.\\
In this Letter we study the NF empirical scaling~\cite{Noro2000} in the
potential energy landscape (PEL) thermodynamic framework, focusing on
particles interacting via square-well (SW) potential of different well-width
$\Delta$, to provide a deeper understanding of the generalized law of
corresponding states.  We first show that for this model the NF-law holds in a
large range of densities (not only close to the liquid-gas critical point) and
correctly reproduce the Baxter behavior in the limit of vanishing $\Delta$.
Then we calculate, with no approximations, the statistical properties of the
PEL~\cite{stillinger,Wales2004,Sciortino2005} and their dependence on the well
width. This study allows us to separate the basin free entropy in two
components, respectively reflecting the exploration of the bond volume
($\Delta-dependent$) and the exploration of the available volume at fixed bond
distance ($\Delta-independent$). This second contribution is the only one
existing in the Baxter model.\\
To generate equilibrated configurations, we perform event-driven molecular
dynamics (EDMD) simulations~\cite{Rapaport1995} for the SW potential, defined
by a potential well of depth $u_0=-1$ and well width $\Delta$.
We investigate values of $\Delta/d$ ranging from $10^{-1}$ to $10^{-3}$, where
$d=1$ is the hard-core particle diameter, chosen as unity of length. Pairs of particles
are considered bonded when their relative distance is between $d$ and $d+\Delta$.
Density is expressed as packing fraction $\phi=\frac{\pi}{6} \rho d^3$, where
$\rho=N/L^3$, $N$ being the number of particles and $L$ the simulation box
size.  
According to the NF law, the thermodynamic of the system is controlled
by $B_2^*$ and $\phi$, at least in the proximity of the critical point. In the
SW case $B_2^*-1 \sim \Delta
e^{u_0/T}$  in the limit of short well-width.\\
Fig.~\ref{fig:NF} shows the potential energy per particle $U/N$ (panel a) and
the compressibility factor $Z=PV/Nk_BT$ (panel b) as a function of $\phi$ for
several $\Delta$ values, at $B_2^*=-0.40$.  For  $\Delta < 0.05$, all
curves collapse on the same curve confirming the validity of NF scaling also
outside the critical region.  Similar behavior is observed for other different
$B_2^*$ values. The value at which the scaling breaks, i.e. $\Delta\sim0.05$,
is in close agreement with the value at which the Baxter solution ceases to be
a valid approximation for the SW model~\cite{Malijevskyetal}.
To confirm that the SW potential approaches the Baxter model in the limit of
vanishing $\Delta$ (and as a confirmation of our ability to equilibrate
configurations at small $\Delta$ values) we compare in Fig.~\ref{fig:NF}c the
recently calculated radial distribution function $g(r)$ of the Baxter
model~\cite{Miller2003} with the the $g(r)$ calculated for a SW system with
$\Delta=10^{-4}$ at $B_2^*=-0.92$. The agreement between the two results is
perfect. The shoulders and the $\delta$-functions indicating the presence of
clusters with well defined structure~\cite{Miller2004} are reproduced both in
their location and their height.\\
The fact that for $\Delta<0.05$ the state of the system is controlled by the
values of $\phi$ and of $\Delta e^{\beta u_0}$, suggests that --- at fixed
$\phi$ --- the configurational part of the partition function $\mathcal{Z}$
must be, in leading order, a function of $\Delta e^{\beta u_0}$.  In the PEL
formalism~\cite{stillinger,Wales2004,Sciortino2005} the phase space is
decomposed into basins of attractions of the local minima of the PEL --- the
so-called inherent structures (IS) --- and the partition function is written as a
sum over all landscape basins, each of them weighted by the local minimum
Boltzmann factor and by the basin free-energy.  In the SW case, we associate
an IS to a specific bonding pattern~\cite{NotaMiller} (thus, the IS energy is
expressed by the total number of bonds $N_b$) and the basin free-energy to the
logarithmic of the (multidimensional) volume $\Omega$ which can be explored in
configuration space without breaking or reforming bonds.  In this respect, the
conceptual operation of thinking $\mathcal{Z}$ as a sum over all possible PEL
basins is equivalent to expressing $\mathcal{Z}$ as a sum over all possible
bonding pattern~\cite{Moreno2005a}.  Since the Boltzmann factor weighting the
probability of each configuration is $e^{\beta u_0 N_b}$, scaling in the
variable $\Delta e^{\beta u_0}$ implies that the number of states $\Omega$
sampled by each fixed bonding pattern must scale with $\Delta$ as
$\Delta^{N_b}$, i.e. each bond is independent and contributes a quantity of
order $\Delta$ to $\Omega$.  This is the hypothesis that we test next.\\
\begin{figure}
\begin{center}
\includegraphics[width=.42\textwidth]{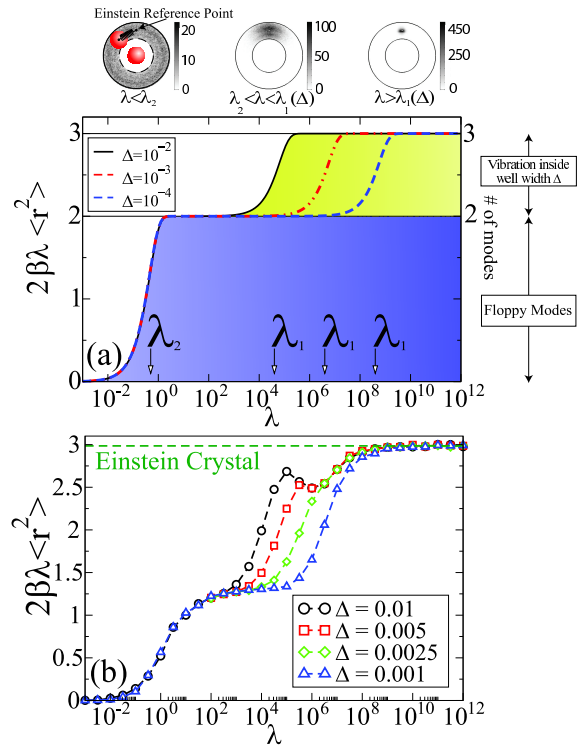}
\end{center}
\caption{(Color online) Top: Mean square displacement of a particle
  harmonically bounded to a reference point and interacting via a SW with a
  particle located in the origin for different $\lambda$ values.  The
  probability distribution of the particle is schematically shown above the
  plot for three representative values of $\lambda$: particle completely
  delocalized (left), particle localized by the bond (center) and particle
  localized by the harmonic force (right). $\lambda_1$ and $ \lambda_2$ are
  discussed in the text.  Bottom: Mean square displacement $2 \beta \lambda
  \langle\sum_i({\bf{r_i}}-\bar{\bf{r}}_i)^2\rangle_{\lambda}/N$ (labelled $2
  \beta \lambda\langle r^2\rangle$ for semplicity in the figure) for a SW
  system with $\phi=0.30$ and $B_2^*=-0.69$ vs. $\lambda$ , for different
  $\Delta$ values. The Einstein limit is shown as a dashed line.}
\label{fig:r2}
\end{figure}
To precisely estimate $\Omega$ we implement a generalization of the
thermodynamic integration technique first used by Ladd and Frenkel to evaluate
the free energy of the hard sphere crystals~\cite{Frenkel2001}.  In this
technique, the system Hamiltonian $H_0$ is complemented by an harmonic term $
\lambda \sum_i(\bf{r_i}-\bar{\bf{r}}_i)^2$ centered around a reference
configuration $(\bar{\bf{r}}_1, \dots, \bar{\bf{r}}_N)$ of strength $\lambda$
acting on each particle.  The essence of this method lies in the possibility
of performing a thermodynamic integration over $\lambda$ from a state of known
free energy to the state of interest ($\lambda=0$).  When $\lambda$ is very
large, particles harmonically oscillate around the reference state and the
system behaves like a collection of $3N$ harmonic oscillators, i.e. an
Einstein solid. If the unperturbed system is identified with a specific
bonding pattern and $H_0$ is defined as zero if the configuration has the
correct bonding pattern and infinity if the bonding pattern has changed (i.e.
if a bond is broken or a new bond is formed), then the result of the
thermodynamic integration provides an exact measure of the basin free energy.
Since the exploration of space within the fixed bond-pattern basin takes place
on a flat energy surface the only contribution to the basin free energy is
entropic.  The basin entropy { per particle} in units of $k_B$,
$\sigma_{\text{basin}}$, can be formally written as
~\cite{Moreno2005a} 
\begin{equation}
\label{eq:ladd2}
\sigma_{\text{basin}} =-\beta
f_{\text{E}}(T,\lambda_\infty)+\int_{0}^{\lambda_\infty}  \lambda \langle\sum_i({\bf{r_i}}-\bar{\bf{r}}_i)^2\rangle_{\lambda} d\ln{\lambda}
\end{equation}
where $f_{\text{E}}(T,\lambda)$ is the free energy of $3N$ harmonic
oscillators coupled by an elastic constant $\lambda$,
$\langle\cdot\rangle_{\lambda}$ an ensemble average at a fixed value of
$\lambda$ and $\lambda_{\infty}$ is any value of $\lambda$ such that the
harmonic contribution in dominant as compared to $H_o$ and thus $\langle
\lambda \sum_i({\bf{r_i}}-\bar{\bf{r}}_i)^2 \rangle_{\lambda} \approx
\frac{3}{2}Nk_BT$.  From a technical point of view we evaluate $\langle
\lambda \sum_i({\bf{r_i}}-\bar{\bf{r}}_i)^2 \rangle_{\lambda}$ for thirty or
more $\lambda$ values between $\lambda_{\infty}$ and $0$ via MC simulations
rejecting all moves which modify the bond-pattern.\\
To guide the interpretation of the numerical results it is helpful to examine
the behavior of two particles bonded by a square-well in three dimensions.
Fixing particle $1$ at the origin (to neglect the trivial center of mass
degrees of freedom) and selecting an Einstein reference site acting on
particle $2$ in an arbitrary position $\bar{\bf{r}}_{\bf{2}} \equiv (x,0,0)$
(with $ d < x < d+\Delta$), the resulting $ \langle
({\bf{r_2}}-\bar{\bf{r}}_{\bf{2}})^2 \rangle_{\lambda} $ is
\begin{equation}
  \lambda\langle        ({\bf{r_2}}-\bar{\bf{r}}_{\bf{2}})^2
  \rangle_{\lambda}  = -\frac{\partial}{\partial \beta} \log   \mathcal{Z}_{\lambda}(\beta),
\end{equation}
where the generating (or partition) function $\mathcal{Z}_{\lambda}(\beta)$ is

\begin{equation}
\mathcal{Z}_{\lambda}(\beta)= 2 \pi   \int_d^{d+\Delta}  \frac{ e^{-
    \beta \lambda (r+x)^2} (e^{- 4  \beta \lambda r x }-1)  }{2 \beta \lambda} r dr
\end{equation}
The resulting $\lambda$ dependence of $2 \beta \lambda \langle
({\bf{r_2}}-\bar{\bf{r}}_{\bf{2}})^2 \rangle_{\lambda} $ is shown in
Fig.~\ref{fig:r2}a for three different values of $\Delta$.  At large $\lambda$
(the harmonic limit), the quantity $2 \beta \lambda \langle
({\bf{r_2}}-\bar{\bf{r}}_{\bf{2}})^2 \rangle_{\lambda} $ goes to three, the
total number of degrees of freedom.
On decreasing $\lambda$, the function shows a two-step decay to zero with an
intermediate plateau at the value two. The two cross-overs (from $3
\rightarrow 2$ and from $2 \rightarrow 0$) are taking place at values of $
\lambda_1 \approx (2/\Delta)^2$ and $ \lambda_2 \approx (2/d)^2$. To interpret
this behavior we recall that for very large $\lambda$ ($\lambda> \lambda_1$)
confinement is provided by the harmonic potential.  For $\lambda_2 < \lambda <
\lambda_1$,  confinement of the harmonic potential has become
larger than $\Delta$ and the bond distance becomes the relevant quantity
controlling the mean square displacement.  For even smaller $\lambda$
($\lambda < \lambda_2$) confinement is provided by the { finite bond
  volume} (a corona of width $\Delta$ and inner surface $4 \pi d^2$).  The
two-step cross-over makes it possible to count the number of modes which are
related to exploration of the bond width and the number of modes which are
related to exploration of space at fixed interparticle distance.  Indeed, the
first cross-over is $\Delta$-dependent while the second one is
$\Delta$-independent. Hence, the FL method allows one not only to evaluate the
total change in entropy but also to count (and separate !) the number of modes
which are related to the exploration of the bond-distance (vibrational modes)
from the modes which are related to the exploration of the volume with rigid
bonds (floppy modes).
\begin{figure}
\begin{center}
\includegraphics[width=.34\textwidth]{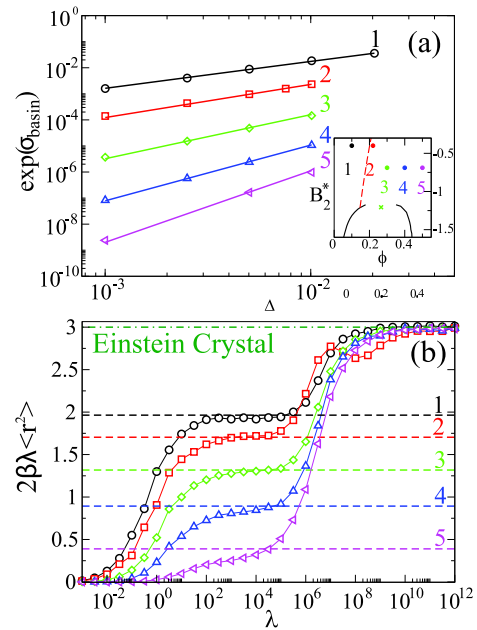}
\end{center}
\caption{(Color online) Top: Number of state sampled
  $\Omega^{1/N}=exp(\sigma_{\text{basin}})$ vs.  $\Delta$ for different state
  points. The straight lines are single parameter fit of the form $A
  \Delta^{n_b}$ where $n_b$ is the number of bonds per particle per particle,
  i.e.  $n_b=N_b/N$.  The inset shows the the location of the corresponding
  state points in the $B_2^*-\phi$ phase diagram. It also show the percolation
  line (dashed line) and the liquid-gas coexistence from
  Ref.\onlinecite{Miller2003} Bottom: Symbols are $2 \beta \lambda
  \langle\sum_i({\bf{r_i}}-\bar{\bf{r}}_i)^2\rangle_{\lambda}/N$ (labelled
  $2 \beta \lambda\langle r^2\rangle$ for semplicity in the figure) vs. $\lambda$
  for $\Delta=10^{-3}$; The dashed horizontal line are $3-n_b$.  Note that the
  intermediate $\lambda$ plateau is well reproduced by the fraction of floppy
  modes $f_f=3-n_b$.}
\label{fig:svib}
\end{figure}
The behaviour of $2 \beta \lambda
\langle\sum_i({\bf{r_i}}-\bar{\bf{r}}_i)^2\rangle_{\lambda}/N$ for a bond
configuration of 200 particles (a typical equilibrium configuration produced
by MD simulation) for different values of $\Delta$ is shown in
Fig.~\ref{fig:r2}b.  The bonding network is identical for all $\Delta$ values.
As in the two-particles example, the shape of the curve shows two parts, one
$\Delta$-independent and one $\Delta$-dependent. As from Eq.~\ref{eq:ladd2},
the area under the $\lambda
\langle\sum_i({\bf{r_i}}-\bar{\bf{r}}_i)^2\rangle_{\lambda}/N$ vs
$\ln{\lambda}$ curve is a measure of $\sigma_{\text{basin}}$, the entropy associated
to the exploration of space at fixed bonding pattern. The $\Delta$-independent
and the $\Delta$-dependent parts of
$\lambda\langle\sum_i(\bf{r_i}-\bar{\bf{r}}_i)^2\rangle_{\lambda}$ give rise to two
different contribution to $\sigma_{\text{basin}}$, which we can identify as the
floppy~\cite{Naumis2005}  $\Delta$-independent --- the dark shaded region in the cartoon in
Fig.~\ref{fig:r2}a --- ($N\sigma_{\text{floppy}
}=\ln{\Omega_{\text{floppy}}}$) and the vibrational $\Delta$-dependent part
--- the light coloured region---
($N\sigma_{\text{vib}}= \ln{\Omega_{\text{vib}}}$).\\
According to the PEL picture of the NF law,
$\Omega_{\text{basin}}^{1/N}=e^{\sigma_{\text{basin}}}$ must be proportional
to $\Delta^{n_b}$, where $n_b \equiv N_b/N$ is the number of bonds per
particle.  This prediction can be put under severe test, since both $\Delta$
and $n_b$ are a-priori known, by comparing the calculated values for
$e^{\sigma_{\text{basin}}}$ with $\Delta^{n_b}$.  This comparison is reported
in Fig.~\ref{fig:svib}a for several typical bonding configurations with
different bonding patterns, extracted from equilibrium simulations at
different $\phi$ and $T$,
encompassing the range $1.03< n_b < 2.61$.  In all studied cases,
$\Omega_{\text{basin}} \sim \Delta^{n_b}$ with extreme accuracy.  The validity
of such relationship suggests that for short-range SW, each bond acts as an
independent unit (so that the vibrational entropy of the bonding pattern
coincides with the sum of all the bond entropies).  In the PEL language, the
NF scaling is an expression of the independence of the bonds.  In this
respect, proper scaling between different potential shapes is predicted when
the vibrational free energy of a bonded pair is chosen as a proper scaling
variable.
\begin{figure}[tbh]
\begin{center}
\includegraphics[width=.31\textwidth]{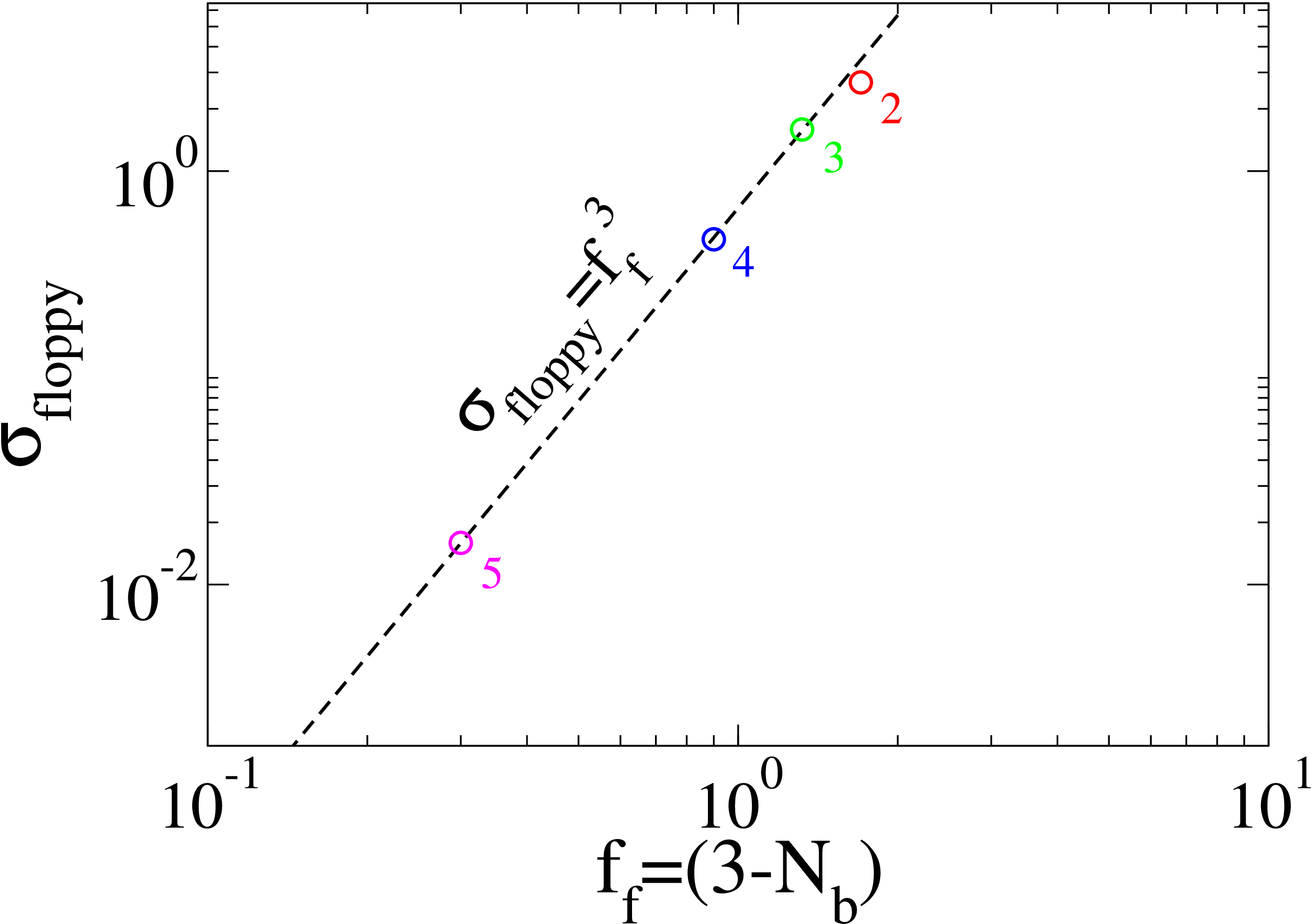}
\end{center}
\caption{(Color online) Entropy of the floppy modes ($\sigma_{\text{floppy}}$)  vs fraction of floppy
  modes ($f_f=3-N_b/N$). The straight line is a power low with exponent
  $3$, i.e. $\sigma_{\text{floppy}}\propto f_f^3$.}
\label{fig4}
\end{figure}
The possibility of separating in a precise way $\Omega_{\text{vib}}$ and
$\Omega_{\text{floppy}}$ (see the cartoon in Fig.\ref{fig:r2}a), allows us to
evaluate also the ($\Delta$-independent) volume in configuration space sampled
by the a specific bonding pattern when all bond distances are fixed.  This
volume corresponds to the free rolling motions of the particles with no
bond-breaking of -forming and it is essentially the basin volume accessible to
the Baxter model.  It is interesting to investigate the dependence of this
quantity on the number of bonds, since one expect that on increasing the
connectivity, the entropy of the floppy modes should decrease. For the
short-range SW under consideration the fraction of floppy modes (the height of
the plateau in $2 \beta \lambda
\langle\sum_i({\bf{r_i}}-\bar{\bf{r}}_i)^2\rangle_{\lambda}/N$) is found to be
$f_f=3-n_b$ (see Fig.~\ref{fig:svib}b), a value consistent with the existence
of $N_b$ independent vibrational degrees of freedom.  Hence, the total floppy
entropy (the area under the curve -- see Eq.~\ref{eq:ladd2}) should vanish
close to the point when $n_b \approx 3$ (i.e. each particle is involved in
average in six bonds).  This expectation is indeed born out in the calculated
$f_f$-dependence of $\sigma_{\text{floppy}}$ shown in Fig.~\ref{fig4}.
Unexpectedly, the floppy entropy goes to zero as a power law
$\sigma_{\text{floppy}}\propto f_f^3$.  The value of the exponent suggests
that not only the number of modes but also the floppy entropy per mode
vanishes with a power-law close to $f_f=0$. It is interesting to observe that
at the state point where $N_b \approx 3N$, the full entropy of the system
would be given only by the logarithm of the number of topologically different
bonding patterns.\\
In conclusion, we have shown that the empirical NF-law of corresponding states
can be explained in the rigorous PEL thermodynamic formalism as arising from
the additive contribution of the bonds to the basin entropy.
An unexpected outcome of this study is a new methodology to separately
evaluate the fraction of floppy modes and their entropy for any specific
bonded configuration, a method which can be used in studies of the rigidity of
hard particle systems~\cite{Thorpe1999}.

We thank D. Frenkel, M. Miller and R. Sear for useful discussions. We also
thank M. Miller for providing us the Baxter $g(r)$ data and C. De Michele for
the EDMD code. We acknowledge support from the Swiss National Science
Foundation (Grant No.  99200021-105382/1) (G.F.), from MIUR-FIRB and Cofin
(F.S.) and from Training Network of the Marie-Curie Programmme of the EU
(MRT-CT-2003-504712).

%
%
%
%
%
%
\bibliography{add,tesi}

\begin{thebibliography}{21}
\bibitem[{Not()}]{NotaMiller}
\bibinfo{note}{Note that our definition of IS does not require a minimization
  procedure. Two configurations related by a path of decreasing energy (i.e.
  which can be connected by moves which requires or zero or a negative change
  of energy are considered belonging to distinct basins.}
\end{thebibliography}
\bibliographystyle{apsrev}

\end{document}